\documentstyle[12pt,aaspp4]{article}

\def\lsim{{{}_{{}_<}^{~}\atop {}^{{}^\sim}_{~}}}
\newcommand{\be}{\begin{equation}}
\newcommand{\ee}{\end{equation}}
\newcommand{\lab}[1]{\label{#1}}
\newcommand{\rr}[1]{~(\ref{#1})}

\begin{document}
\title{Clump Giant Distance to the Magellanic Clouds and Anomalous Colors
in the Galactic Bulge}
 \author{Piotr Popowski}
\affil{Institute of Geophysics and Planetary Physics, L-413\\ Lawrence Livermore
National Laboratory, University of California\\ P.O.Box 808, Livermore, CA 94551, USA.\\ E-mail: popowski@igpp.llnl.gov.}

\begin{abstract}
I demonstrate that the two unexpected results in the local Universe: 
1) anomalous intrinsic
$(V-I)_0$ colors of the clump giants and RR Lyrae stars in the Galactic 
center, and
2) very short distances to the Magellanic Clouds (LMC, SMC) as inferred from 
clump giants, are connected with each other.
The $(V-I)_0$ anomaly is partially resolved by using the photometry from
the phase-II of the Optical Gravitational Lensing Experiment (OGLE) rather
than phase-I. The need for V- or I-magnitude-based change in the bulge 
$(V-I)_0$ is one option to explain the remaining color discrepancy. Such 
change may originate in a coefficient of selective extinction $A_V/E(V-I)$ 
smaller than typically assumed. Application of the $(V-I)_0$ correction 
(independent of its source) doubles the slope of the absolute magnitude -- 
metallicity relation for clump giants, so that $M_I(RC) = -0.23 + 0.19
\, {\rm [Fe/H]}$. 
Consequently, the estimates
of the clump distances to the LMC and SMC are affected. Udalski's 
(1998c) distance modulus of $\mu_{LMC} = 18.18 \pm 0.06$ increases
to $\mu_{LMC}= 18.27 \pm 0.07$. 
The distance modulus to the SMC increases by 0.12 to $\mu_{SMC}=18.77
\pm 0.08$.
I argue that a more comprehensive assessment of the metallicity effect
on $M_I(RC)$ is needed.
  
Subject Headings: distance scale --- dust, extinction --- Galaxy: center --- Magellanic Clouds --- stars: horizontal-branch
\end{abstract}

\section{Introduction}
Most of the extragalactic distance scale is tied to the LMC, and so the 
distance to the LMC ($d_{LMC}$) influences the Hubble constant, $H_0$.
For many years now there has been a division between the so called ``short''
and ``long'' distance scales to the LMC. Currently, the measured values of
$d_{LMC}$ span a range of over 25\% (see e.g., Feast \& Catchpole 1997; 
Stanek, Zaritsky, \& Harris 1998).
Paczy\'{n}ski \& Stanek (1998) pointed out that red clump giants
should constitute an accurate
distance indicator. Udalski et al. (1998a) and Stanek et al. (1998) applied 
the clump method and found a very short distance to the LMC ($\mu_{LMC} 
\approx 18.1$). In response, Cole (1998) and Girardi et al. (1998) suggested 
that clump giants are not standard candles and that their
absolute $I$ magnitudes, $M_I(RC)$, depend on the metallicity and age of the
population. Udalski (1998b, 1998c) countered this criticism by showing that 
the metallicity dependence is at a low level of about $0.1$ mag/dex, and that 
the $M_I(RC)$ is approximately constant for cluster ages between 2 and 10 Gyr.
Stanek et al. (1999) and Udalski (1999) found a moderate slope of the 
$M_I(RC)$ -- [Fe/H] relation of 0.15 mag/dex.
The only clump determination, which resulted in a truly long $d_{LMC}$ was a 
study of the field around supernova SN 1987A by Romaniello et al. (1999).
However, they assumed a bright $M_I(RC)$ from theoretical models and, 
additionally, the use of the vicinity of SN 1987A may not be the most 
fortunate choice (Udalski 1999).

The value of $M_I(RC)$ in different stellar systems is a major systematic 
uncertainty in the clump method.
It is very hard to prove the standard character of a candle's luminosity. 
However, it should be
possible to check whether other stellar characteristics of a candle behave in
a predictable fashion. Therefore, in \S 2 I discuss the $(V-I)_0$ colors
of the clump giants and RR Lyrae stars in the Galactic bulge. After making
photometric corrections, I argue that the remaining color discrepancy 
between the Baade's Window and local stars
might have been caused by an overestimated coefficient of selective 
extinction. Using corrected colors, in \S 3 I derive a new
$M_I(RC)$ -- [Fe/H] relation for red clump stars and show its substantial
impact on the distances to the Magellanic Clouds. I summarize the results
in \S 4.

\section{Mystery of anomalous colors in the Galactic bulge}

Paczy\'{n}ski (1998) tried to explain why the clump giants
in the Baade's Window have $(V-I)_0$ colors which are approximately $0.2$
magnitudes redder than in the solar neighborhood (Paczy\'{n}ski \& 
Stanek 1998). Paczy\'{n}ski (1998) suggested super-solar metallicities of the 
Galactic bulge stars as a possible solution. However, there is 
a spectroscopic evidence (see Minniti et al. 1995) that the average 
metallicity of the bulge is [Fe/H] $ \in (-0.3, 0.0)$.
Stutz, Popowski \& Gould (1999) found a corresponding
effect for the Baade's Window RR Lyrae stars, which have $(V-I)_0$ redder by
about 0.17 than their local counterparts (Fig.\ 1).

\placefigure{figure1}

A similar size of the color shift in RR Lyrae stars and clump giants
suggests a common origin of this effect. Does there exists any physical 
mechanism that could be responsible for such behavior?
The bulge RR Lyrae stars and clump giants both burn Helium in their cores, 
but the similarities end here. 
RR Lyrae stars pulsate, clump giants do not. RR Lyrae stars are metal-poor,
clump giants are metal-rich.
RR Lyrae stars are likely to be a part of an axisymmetric stellar halo (e.g., 
Minniti 1996; Alcock et al. 1998a), whereas clump giants form a bar (e.g., 
Stanek et al. 1994; Ng et al. 1996).
For RR Lyrae stars, Stutz et al. (1999) suggested that their very red 
$(V-I)_0$ might have resulted from an unusual abundance of $\alpha$- elements.
Why should a clump population which emerged in a different formation
process share the same property?

The solutions to the anomalous colors proposed by Paczy\'{n}ski (1998)
and Stutz et al. (1999) are not impossible but are rather unlikely.
Alternatively, the effect might be unrelated to the physics of those stars.
The investigated bulge RR Lyrae stars and clump giants share two things
in common. First, photometry of both types of stars comes from the OGLE, 
phase-I, project. Indeed, Paczy\'{n}ski et al. (1999) showed that the
OGLE-I V-magnitudes are 0.021 mag fainter and I-magnitudes 0.035 mag brighter
than the better calibrated OGLE-II magnitudes. Therefore, the correct $(V-I)$ 
colors should be 0.056 bluer. Additionally, the new $(V-I)_0$ from the more 
homogeneous Baade's Window clump is bluer than Paczy\'{n}ski's \& Stanek's 
(1998) color even when reduced to OGLE-I calibration\footnote{Udalski's
(1998b) data for the LMC, SMC, and Carina galaxy come from OGLE-II and 
therefore do not require any additional adjustment.}. 
When the new OGLE-II photometry reported by Paczy\'{n}ski et al. (1999) is 
used, the $(V-I)_0$ anomaly shrinks and the remaining unexplained shift 
amounts to $\sim 0.11$ both for the RR Lyrae stars and clump giants.

Second, Paczy\'{n}ski \& Stanek (1998) and Stutz et al. (1999) use the same
extinction map (Stanek 1996) and the same coefficient of conversion from 
visual extinction $A_V$ to color excess $E(V-I)$. The absolute values of 
$A_V$s are likely approximately correct (see equation \ref{avadjusted}) 
because the zero point of the 
extinction map was determined from the $(V-K)$ color and $A_V/E(V-K)$ is 
very close to 1 (Gould, Popowski, \& Terndrup 1998; Alcock et al. 1998b).
However, $R_{VI}=A_V/E(V-I)$ is not as secure and has a pronounced effect
on the obtained color.

Most of the current studies of the Galactic bulge use $R_{VI} = 2.5$. 
If a true $R_{VI}$ towards Baade's Window
equals $\alpha$ instead, then the adjusted Stanek's (1996) V-band extinction,
will be\footnote{Equation\rr{avadjusted} implicitly assumes that 
{\em differential} $(V-I)$ colors from Stanek (1996) are correct. Whether
it is the case is an open question.}:
\be
A_{V,{\rm adjusted}} = \frac{\alpha}{2.5} \left( A_V - A_{V, \mbox{\scriptsize 0-point}} \right) + A_{V, \mbox{\scriptsize 0-point}}, \lab{avadjusted}
\ee
where most of the extinction, namely $A_{V, \mbox{\scriptsize 0-point}}$ is
excluded from the adjustment because it has been determined based on
$K$-magnitudes\footnote{To the zero-th order, a shape of the extinction
curve is governed by one parameter, so that the change in $R_{VI}$ will
affect $A_V/E(V-K)$. In effect, Gould et al. (1998) and Alcock et al. (1998)
might have overestimated (underestimated) $A_{V, \mbox{\scriptsize 0-point}}$
by a few percent if $\alpha < 2.5$ ($\alpha > 2.5$). This change would
propagate to equations\rr{avadjusted} -\rr{alpha} in several
correlated ways. Thus, strictly speaking a value of $\alpha$ is 
additionally a function of $A_V/E(V-K)$, which changes with $\alpha$. 
Therefore, solving this problem exactly requires iteration.
I neglect this complication in further considerations.}.
The adjustment to the color, which follows from equation\rr{avadjusted} is:
\be
\Delta (V-I)_0 =  \frac{1}{2.5} A_V - \frac{1}{\alpha} A_{V,{\rm adjusted}} = \frac{\alpha - 2.5}{2.5\,\alpha} A_{V, \mbox{\scriptsize 0-point}}. \\
\lab{deltavmi} 
\ee
Therefore, for a color shift $\Delta (V-I)_0$ , one expects:
\be
\alpha = \frac{2.5 \, A_{V, \mbox{\scriptsize 0-point}}}{A_{V, \mbox{\scriptsize 0-point}} - 2.5 \, \Delta (V-I)_0}. \lab{alpha}
\ee
Using $\Delta (V-I)_0 \approx -0.11$ as required to resolve the color conflict
in Baade's Window and $A_{V, \mbox{\scriptsize 0-point}}=1.37$ (Gould et al.
1998; Alcock et al. 1998b), I find $\alpha \approx 2.1$ (Fig. 2).
This $R_{VI} = 2.1$ is certainly low, but not unreasonably so. 
Szomoru \& Guhathakurta (1999) find that cirrus clouds in the Galaxy have 
extinctions consistent with $A_V/E(B-V) \lsim 2$, which is more extreme than 
the change suggested here. If the extinction towards Baade's Window is in 
part provided by the cirrus clouds, then the low $R_{VI}$ would be expected 
rather than surprising.

\placefigure{figure2}

The value and variation of $R_{VI}$ was thoroughly investigated by
Wo\'{z}niak \& Stanek (1996). The essence of the Wo\'{z}niak \& Stanek (1996) 
method to determine differential extinction is an assumption that regions
of the sky with a lower surface density of stars have higher extinction.
Wo\'{z}niak \& Stanek (1996) used clump giants to convert a certain density 
of stars to an amount of visual extinction.
To make a calibration procedure
completely unbiased would require, among other things, that clump giants were 
selected without any assumption about $R_{VI}$; that absolute
$V$-magnitudes of clump giants, $M_V(RC)$, do not depend on their color 
[here $(V-I)_0$]; and that reddened and unreddened clump giants be drawn from 
the same parent population. None of these is true.
A color-magnitude diagram (CMD) for dense Galactic fields does not allow one
to unambiguously distinguish clump giants from other stars. Different parts
of an intrinsically clean CMD overlap due to differential 
reddening and a range of stellar distances. Therefore, the selection
of clump giants must involve some assumptions about $R_{VI}$.
Wo\'{z}niak \& Stanek (1996) adopt $R_{VI} = 2.6$.
This procedure tends to 
bias the derived relation toward this predefined slope. Wo\'{z}niak \& Stanek 
(1996) were fully aware of this effect, and they performed a number of 
simulations, which are summarized in their Figure 4. In brief, in the range 
$2.1 < R_{VI} < 3.1 $, the bias scales as $\delta R_{VI} \sim 0.4 \, 
(2.6 - R_{VI})$ and so may become very substantial for a low or high $R_{VI}$.
In particular, if the true $R_{VI}=2.1$, Wo\'{z}niak and Stanek (1996) would
find $R_{VI}=2.3$. Therefore, this effect alone could account for half
of the difference between the required and measured $R_{VI}$.
 
The intrinsic characteristics of the bulge clump stars are unknown, but
I will assume they resemble the clump measured by Hipparcos (European
Space Agency 1997). The fit to the local clump giants selected by 
Paczy\'{n}ski \& Stanek (1998) gives $M_V(RC) \propto 0.4(V-I)_0$. 
Therefore, the structure of the local clump itself acts similarly to 
extinction with $R_{VI} = 0.4$. In an ideal case, when the CMD locations
of the entire clump populations in different fields are compared, the 
$M_V(RC) - (V-I)_0$ dependence should not matter. However, when combined with 
the actual extinction and additionally influenced by the completeness 
function of a survey, this effect may additionally bias the value of $R_{VI}$.

Because the smaller selective extinction coefficient is not excluded by the 
current studies, one can assume $R_{VI}=2.1$ to match the $(V-I)_0$ colors of 
the bulge with the ones in the solar neighborhood. The color is a weak 
function of [Fe/H], so this procedure is justified because the [Fe/H] 
of the bulge and solar neighborhood are similar. This change in $R_{VI}$ will 
decrease the I-mag extinction, $A_I$, by 0.11 mag. Therefore, the clump 
distance to the Galactic center would increase by the same amount.

\section{Recalibration of the clump}

What is the bearing of the bulge results on the distance to the LMC?
Let $\Delta$ indicate the difference between the mean dereddened I-magnitude
of clump giants and the derredened V-magnitude of RR Lyrae stars at the 
metallicity of RR Lyrae stars in the Galactic bulge. When monitored in 
several stellar systems with different clump metallicities, the variable 
$\Delta$, introduced by 
Udalski (1998b), allows one to calibrate the $M_I(RC)$ - [Fe/H] relation with 
respect to the baseline provided by RR Lyrae stars.
The better photometry from Paczy\'{n}ski et al. (1999) and a possible
modification of $R_{VI}$ influence the value of $\Delta$ at the Galactic 
center ($\Delta_{BW}$). It is important to note that one will face the same 
type of adjustment to $\Delta_{BW}$ whenever the anomalous colors in the 
Baade's Window are resolved at the expense of the modification of 
V- or I-magnitudes.
That is, the modification of $R_{VI}$ is not a necessary condition! It 
is simply one of the options. As a result of the change in $\Delta_{BW}$, 
the $M_I(RC)$ - [Fe/H] relation for clump giants changes. 
Moreover, $\mu_{LMC}$ and $\mu_{SMC}$ will change as well because 
the $M_I(RC)$ -- [Fe/H] relation is used to obtain the clump distances to the 
Magellanic Clouds. 

Here, I will modify Udalski's (1998b) $\Delta$ versus [Fe/H] plot
and derive a new $M_I(RC)$ - [Fe/H] relation consistent with the new
data and considerations from \S 2.
I construct the Udalski (1998b) plot using his original points modified
in the following way:\\[0.1cm]
--- To match the change in $(V-I)_0$, I modify $\Delta_{BW}$ by 0.17 mags 
(a combined change from photometry and some other, yet unrecognized, source, 
e.g., selective extinction coefficient).\\[0.1cm]
--- I modify the [Fe/H] of the Baade's Window clump giants, so that 
${\rm [Fe/H]}=0.0$ (see e.g., Minniti et al. 1995 for a review on the bulge
metallicity).
\\[0.1cm]
The possible improvement to the above procedure would be
a construction of Udalski's (1998b) diagram based on clump giants in the LMC 
and SMC clusters, which would reduce the uncertainties associated with the
reddening to the field stars. This more complex treatment is beyond the
scope of this paper.

I make a linear fit to the $\Delta$ -- [Fe/H] relation.
I assume that a total error in dependent variable $\Delta$
for the $i$-th point, $\sigma_{total,i}$, can be expressed as
$ {\sigma}_{total,i}^2 = {\sigma}_{\Delta,i}^2 + {\left( \left. \frac{d\Delta}{d {\rm [Fe/H]}} \right|_{{\rm [Fe/H]}_i} {\sigma}_{{\rm [Fe/H]},i} \right)}^2,
$
where ${\sigma}_{\Delta,i}$ and ${\sigma}_{{\rm [Fe/H]},i}$ are the 
individual point errors in $\Delta$ and metallicity, respectively.
As a result, I obtain:
\be
M_I(RC) = (-0.36 \pm 0.03) + (0.19 \pm 0.05) \, ({\rm [Fe/H]}+0.66) 
\lab{clumpmag}
\ee
Equation\rr{clumpmag} is expressed in the form with uncorrelated errors
and normalized
to the local Hipparcos (European Space Agency 1997) result of $M_I(RC) = -0.23 
\pm 0.02$ at [Fe/H] $= 0.0$ reported by Stanek \& Garnavich (1998).
Equation\rr{clumpmag} is a good fit to the data with a $\chi^2/d.o.f = 3.17/2$
(Fig. 3).
The slope from equation\rr{clumpmag} is 0.10 mag/dex steeper than the
one given by Udalski (1998b). However, it agrees well with the slope of 
$0.15 \pm 0.05$ based on the spectroscopic data for the local clump 
(Udalski 1999).
Also, the slope from equation\rr{clumpmag} is in a good agreement with
the theoretical models (e.g., Cole 1998) and leads to $M_{I,LMC}(RC)=-0.35$
in the LMC field and $M_{I,LMC}(RC)=-0.39$ for the clusters (close to the 
values suggested by Girardi et al. 1998).

\placefigure{figure3}

\section{Summary}

I demonstrated that the correction of the bulge $(V-I)_0$ anomaly
has a pronounced effect on the slope of the absolute magnitude -- metallicity 
relation for clump 
giants. Introducing color correction to the original Udalski (1998b)
diagram, I find
$M_I(RC) = -0.23 + 0.19 \, {\rm [Fe/H]}$. 
Consequently, the Udalski's (1998c) distance modulus of $\mu_{LMC} = 18.18 
\pm 0.06$ is increased to $\mu_{LMC}= 18.27 \pm 0.07$. 
The distance modulus to the SMC increases from $\mu_{SMC}=18.65 \pm 0.08$
to $\mu_{SMC}=18.77 \pm 0.08$.

Even though my approach in this paper is only qualitative, there are
two important characteristics of this study:\\[0.1cm]
1) The calibration of $M_I(RC)$ -- [Fe/H] relation, has been based on the 
homogeneous set of the OGLE-II photometry. Therefore, no corrections
due to the use of different telescopes, instruments and reduction procedures
are required. Unfortunately, this makes the above calibration vulnerable to
unrecognized systematic problems of the OGLE photometry.\\[0.1cm]
2) The $M_I(RC)$ value has been derived based on observational data and
not simply picked from a family of possible theoretical models of 
stellar evolution.\\[0.1cm]
Romaniello et al. (1999) provide an independent source of
clump photometry in the LMC, but due to the importance of photometric 
homogeneity I am not able to use their data in a way consistent with 
the rest of my analysis.
With reference to point 2), it is crucial to note that observationally 
calibrated $M_I(RC)$ is not subject to the modeling uncertainties which 
affect the Romaniello et al. (1999) distance to the LMC.
However, my calibration is only as good as the assumptions and data that
enter the analysis. 
Reddening corrections to the original Udalski's (1998b) diagram,
which is partly based on the field stars in the LMC and SMC, may be needed.
Therefore, a more comprehensive study of the metallicity effect on $M_I(RC)$ 
is necessary. Udalski's (1999) determination based on the local clump is an 
important step toward establishing a reliable $M_I(RC)$ -- [Fe/H] relation.

\acknowledgments

Andy Becker deserves my special thanks for many stimulating discussions
about the extinction issues in the Galactic bulge. I am deeply grateful
to Andy Gould for his very careful reading of the original version of this
paper and a number of insightful remarks. I also would like to thank Kem
Cook for his valuable comments and discussions.
Work performed at the LLNL is supported by the DOE under contract 
W7405-ENG-48.

\clearpage

\figcaption[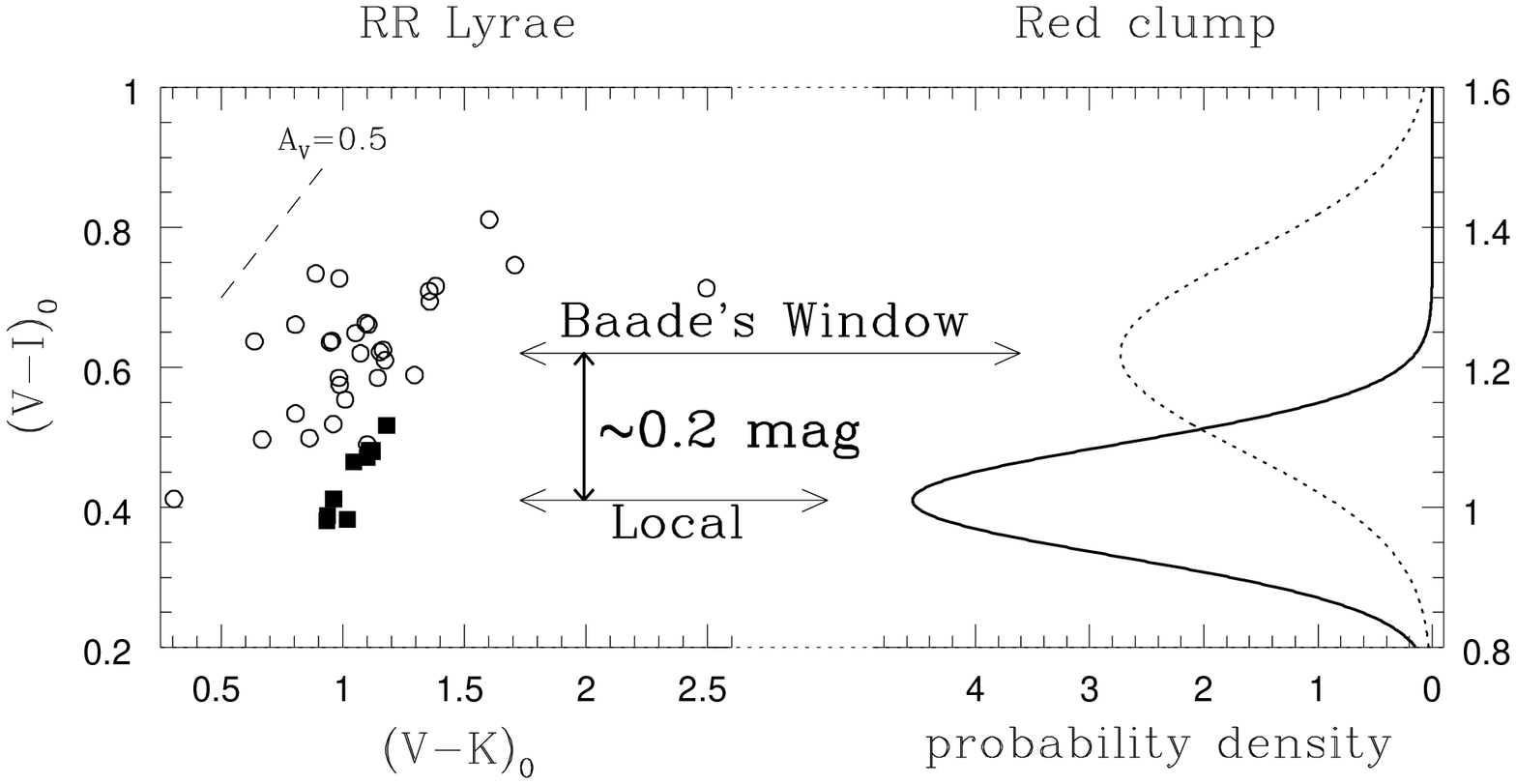]
{Left part presents $(V-I)_0$ colors of the Baade's Window (open circles)
and local (full squares) RR Lyrae stars as a function of $(V-K)_0$. The 
$(V-I)_0$ shift between the two groups is of order of 0.17 mag. The standard 
extinction line (dashed) is parallel to the stellar lines and so will not
cure the anomaly.
The right panel shows a similar situation for clump giants. 
The vertical $(V-I)_0$ color axis should be treated as an abscissa. The two
Gaussians are Paczy\'{n}ski's \& Stanek's (1998) fits to the number densities 
of clump giants as a function of color: in the solar neighborhood 
(thick solid line) and in the Baade's Window (thin dotted line). The $(V-I)_0$ 
difference between the maxima of the two clump groups is 0.21.
\label{figure1}}

\figcaption[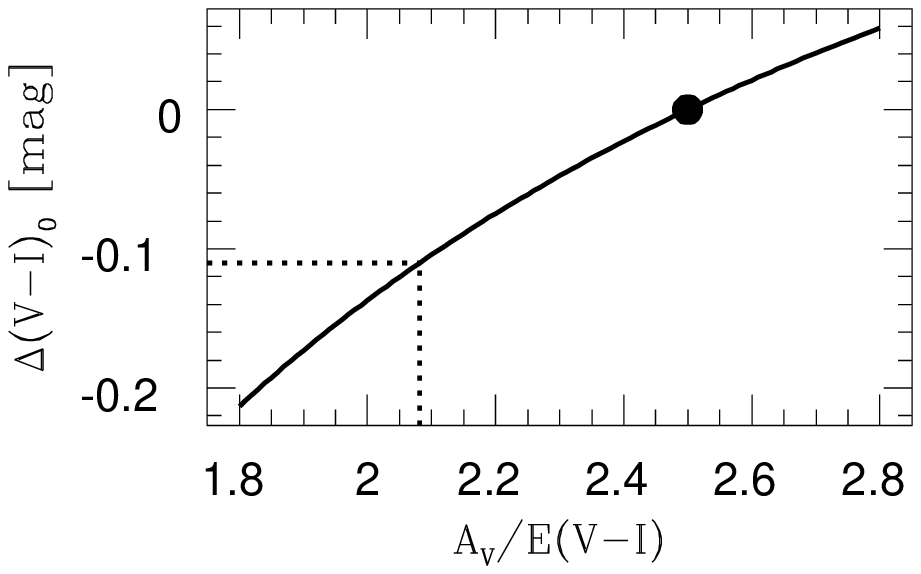]
{The expected adjustment to the $(V-I)_0$ as a function of the selective
extinction coefficient $R_{VI}$. The full dot marks the standard extinction.
The dotted lines indicate the new point which would explain
the entire color anomaly seen in the Baade's Window.
\label{figure2}}

\figcaption[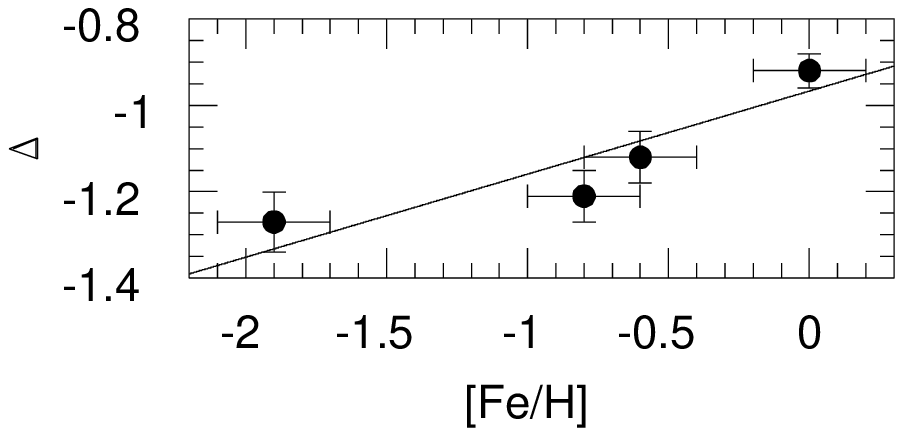]
{Udalski's (1998b) diagram showing $\Delta$ versus [Fe/H].
In order of increasing [Fe/H], the data points correspond to the Carina,
SMC, LMC, and Baade's Window. 
The V-magnitude adjustment to RR Lyrae stars is limited to the OGLE-II-based
correction in Baade's Window. The possible changes to $\Delta$ that move 
{\em all} the points on the diagram by the same amount can be ignored because
they do not influence the calibration.
\label{figure3}}

\end{document}